\documentclass[useAMS,usenatbib]{mnras}

\usepackage{graphicx}
\usepackage{textcomp}
\usepackage{amssymb}\usepackage{hyperref}
\usepackage{amsmath,alltt}           
                          \usepackage{tikz}
\usetikzlibrary{shapes.geometric}

\usepackage[T1]{fontenc}
\usepackage[utf8]{inputenc}
\usepackage{multirow}
\usepackage{rotating}
\usepackage{lscape}
\usepackage{times}
\usepackage{pdflscape}

\usepackage{stmaryrd}

\usepackage{colortbl}

\newcommand\lsim{\mathrel{\rlap{\lower4pt\hbox{\hskip1pt$\sim$}}
        \raise1pt\hbox{$<$}}}
\newcommand\gsim{\mathrel{\rlap{\lower4pt\hbox{\hskip1pt$\sim$}}
        \raise1pt\hbox{$>$}}}
\newcommand{\be}{\begin{equation}}
\newcommand{\ba}{\begin{eqnarray}}
\newcommand{\ee}{\end{equation}}
\newcommand{\ea}{\end{eqnarray}}

\title[Small-N Collisional Dynamics V]{Small-N Collisional Dynamics V:  Beyond the Realm of Not-So-Small-N}

\author[Barrera et al. ]{Carlos Barrera$^1$, Nathan W. C. Leigh$^{1,2}$, Basti\'an Reinoso$^3$, Amelia M. Stutz$^{1,4}$, \and Dominik Schleicher$^{1}$\\
$^{1}$Departamento de Astronom\'ia, Facultad de Ciencias F\'isicas y Matem\'aticas, Universidad de Concepci\'on, Concepci\'on, Chile \\
$^{2}$Department of Astrophysics, American Museum of Natural History, New York, NY 10024, USA\\
$^{3}$Universit\"at Heidelberg, Zentrum f\"ur Astronomie, Institut f\"ur Theoretische Astrophysik, Albert-Ueberle-Str. 2, 69120 Heidelberg, Germany\\
$^{4}$Max-Planck-Institute for Astronomy, Konigstuhl 17, 69117 Heidelberg, Germany}

\begin{document}

\date{Accepted. Received; in original form}

\pagerange{\pageref{firstpage}--\pageref{lastpage}} \pubyear{2008}

\maketitle

\label{firstpage}

\begin{abstract}
Direct collisions between finite-sized particles occur commonly in many areas of astrophysics.  Such collisions are typically mediated by chaotic, bound gravitational interactions involving small numbers of particles.  An important application is stellar collisions, which occur commonly in dense star clusters, and their relevance for the formation of various types of stellar exotica.  In this paper, we return to our study of the collision rates and probabilities during small-number chaotic gravitational interactions ($N$ $\lesssim$ 10), moving beyond the small-number particle limit and into the realm of larger particle numbers ($N$ $\gtrsim$ 10$^3$) to test the extent of validity of our analytic model as a function of the particle properties and the number of interacting particles.  This is done using direct $N$-body simulations of stellar collisions in dense star clusters, by varying the relative numbers of particles with different particle masses and radii.  We compute the predicted rate of collisions using the mean free path approximation, adopting the point-particle limit and using the sticky-star approximation as our collision criterion. We evaluate its efficacy in the regime where gravitational-focusing is important by comparing the theoretical rates to  numerical simulations.
Using the tools developed in previous papers in this series, in particular Collision Rate Diagrams, we illustrate that our predicted and simulated rates are in excellent agreement, typically consistent with each other to within one standard deviation.

\end{abstract}

\begin{keywords}
Key words: gravitation – binaries (including multiple): close – globular clusters: general – stars: kinematics and dynamics – scattering – methods: analytical.
\end{keywords}

\section{Introduction}

Direct collisions between finite-sized particles occur commonly in many areas of astrophysics.  Examples include stellar collisions in dense star clusters \citep{leonard89,leigh11,Fujii13}, collisions of protostars in protoclusters \citep{Klessen11,Moeckel11}, collisions between star clusters in galaxies \citep{leigh20}, galaxy-galaxy collisions in galaxy groups or clusters \citep[e.g.][]{silk19,leigh20}, and even the build-up of planetessimals in protoplanetary disks \citep[e.g.][]{goldreich04}. For instance, \citet{Martin11} proposed a mechanism for disk formation after two low-mass stars collide, which could be a possible origin for hot Jupiter planets.

An important application is direct stellar collisions, which occur commonly in dense star clusters.  These are thought to be important for the formation of various types of stellar exotica.  Such direct stellar collisions are typically mediated by chaotic, bound gravitational interactions involving small numbers of particles.  In clusters, this can occur when single, binary and even triple star systems undergo direct interactions \citep{leigh11,leigh13}.  These types of small-number interactions can mediate exchange interactions that swap compact objects into binaries \citep[e.g.][]{verbunt87}, and even direct stellar collisions during the interactions.  The latter mechanism is thought to be responsible for the production of different types of curious binary star systems, including low-mass x-ray binaries \citep[e.g.][]{pooley06} and millisecond pulsars \citep{verbunt87}, whereas the former are thought to contribute to the formation of blue stragglers via collisions involving main-sequence stars \citep[e.g.][]{leigh07,knigge09,leigh11b,leigh13}, the build-up of massive stars \citep[e.g.][]{Bonnell98,PZ99,Oh12,Oh18} which in turn may produce massive black holes \citep[e.g.][]{PZ04}, seeding the growth of super-massive black holes \citep[SMBHs;][]{Katz15,Sakurai17,Boekholt18,Tagawa20}, and so on.

In the regime of very high central densities, direct collisions between single stars can also occur at significant rates.  For example, \citet{Bonnell98} showed that massive stars can form in the cores of dense young star clusters by accretion-induced collisions.  Many other works have considered the formation of massive stars, typically in the regime of runaway collisions and in the context of the formation of the first massive black holes to form in the Universe \citep[e.g.][]{Devecchi09,Katz15,Sakurai17,Reinoso18,Reinoso20}. It has also been suggested that stellar collisions may be important in nuclear stellar clusters, where a massive black hole would be the inevitable outcome for very dense systems \citep[e.g.][and references therein]{leigh13b,leigh16a,Escala20}. Most such studies use $N$-body simulations to study the relevant physics, but these tend to come along with significant computational expense, limiting the total number of simulations that can be performed and hence the range of initial conditions that can be explored, not to mention the inclusion of more sophisticated physics such as mass loss during stellar collisions \citep{Alister20}, or even simply accurately modelling very massive (10$^{6-8}$~M$_\odot$) stellar systems.  Even with state of the art GPU-accelerated $N$-body codes including hybrid parallelization strategies \citep{Wang15}, such massive systems remain extremely difficult to simulate. 
We are sorely lacking the confirmation of analytic models that might help to
alleviate the initial conditions problem and facilitate a faster and more
extensive exploration of the relevant parameter space.  This is our focus in
this study. 

Analytic estimates for the rate of direct stellar collisions in the literature typically rely on the mean free path approximation, or $n\sigma_{\rm coll}v$ based calculations, see \citet{leonard89} and \citet{leigh11} for examples used directly in this paper, where $n$ is the local stellar number density, $v$ is the relative velocity at infinity of the interloping star in the frame of reference of the target object and $\sigma_{\rm coll}$ is the cross-section for collision.  In dense star clusters, the relative velocities between stars can be sufficiently low relative to their characteristic masses and sizes, that gravitational-focusing is non-negligible.  In these cases, the gravitationally-focused cross-section $\sigma_{\rm coll} =$ $\sigma_{\rm gf}$ for collision is adopted, instead of the geometric cross-section for collision, i.e., $\pi$($R_{\rm 1}$ $+$ $R_{\rm 2}$)/2, where $R_{\rm i}$ denotes the physical radius of star $i$ and we adopt the sticky sphere approximation to define collisions as occurring when the stellar radii overlap in both time and space.  For more details about the specific rate estimates used in this paper, we refer the reader to the more detailed derivations provided in section 2.1 in \citet{leigh17}.

In Paper I of this series \citep{leigh12}, we studied how the number of interacting particles affects the probability of collision, finding a connection between the mean free path approximation and the binomial theorem. For identical particles (and a well-defined total encounter energy and angular momentum), the collision probability scales roughly as $N^2$, where $N$ is the number of interacting particles. The physical origin of this $N$-dependence comes from the binomial theorem; the number of ways of selecting any pair of particles from a set of $N$ identical particles is \citep{leigh12}:
\begin{equation}
\label{eqn:binom}
\binom{N}{2}=\frac{N(N-1)}{2}.
\end{equation}

In Paper II \citep{leigh15} we found that, for the types of small-number interactions expected to occur in actual star clusters, the collision probability is directly proportional to the collisional cross-section, at least for the case of (near-)identical mass particles having different radii. 

In Paper III \citep{leigh17}, we derived analytic formulae for the time-scales or rates for different collision scenarios to occur.  We compared the results to numerical scattering simulations of binary-binary interactions.  By assuming either purely radial or purely tangential motions for the particles, we showed that the simulated relative collision probabilities are bounded by these corresponding analytic predictions. In the purely radial limit, by comparing to our simulations, we further showed that our analytic time-scales provide excellent order-of-magnitude estimates for the mean time-scales for direct collision scenarios involving different particle types.

In Paper IV \citep{leigh18b}, we studied the probabilities for different collision scenarios to occur, assuming in our simulations various combinations of particle masses and radii. We went on to test the framework underlying our model in the Newtonian limit and, in particular, our newly introduced toolkit called Collision Rate Diagrams.  Our method is founded on a combinatorics-based backbone designed to calculate the time-scales or rates for direct collisions to occur during chaotic gravitational interactions involving finite-sized particles with different (but comparable) combinations of masses and radii.

In this paper, the fifth in the series, we continue our study of combinatorics in chaotic Newtonian dynamics. 
We return to our study of the rate of collisions, moving beyond the small-number particle limit (i.e., $N$ $\lesssim$ 10) and into the realm of larger particle numbers (i.e., $N$ $\gtrsim$ 10$^3$) \citep{leigh12,leigh15,leigh17,leigh18b}.  This is done using direct $N$-body simulations performed using the {\small NBODY6} code \citep{Aarseth99}.  Using the tools developed in previous papers in this series, in particular Collision Rate Diagrams, we test the robustness of the mean free path approximation in the large-$N$ limit.  We bridge our previous results focused on few-body interactions and the small-number particle limit to larger particle numbers, in the regime where gravitational-focusing is important.  To this end, we vary the relative numbers of particles with different combinations of masses and radii, and compare the rates obtained from our numerical simulations to the analytically predicted rates.  

In Section~\ref{model}, we begin by revisiting our analytic model to compute relative collision rates for different particle types, and present the suites of numerical N-body simulations used to test the robustness of our model.  In Section~\ref{results}, we present our results, first assuming all identical particles and then adopting different particle types (i.e., having different masses and radii).  In Section~\ref{discussion}, we discuss our results in terms of the quality of the agreement between the analytic and simulated rates, along with the implications of our results for improving the model in future work.  
We summarize our key conclusions in Section~\ref{summary}.

\section{Methods} \label{model}

In this section, we begin with a review of the rates for direct single-single collisions in dense stellar environments, as derived and presented in \citet{leonard89} and \citet{leigh11}.  We go on to present the N-body simulations used to test the analytic collision rates, and describe our assumed structural parameters and initial conditions for the simulations.

\subsection{A simple analytic model} \label{simple}

Consider a self-gravitating, virialized system of point particles interacting via Newtonian gravity.
We assume a total of $N_{\rm obj} = \sum_{\rm i}N_{\rm i}$ , where i $=$ A, B, C and D represents four different groups of particles.  Hence, $N_{i}$

is the number of isolated single particles with mass m$_{\rm i}$ and radius R$_{\rm i}$.

The system forms a gravitationally-bound, approximately spherical cluster, which we assume obeys a Plummer density $\rho(r)$ profile with central density $\rho_{\rm 0}$ and central velocity dispersion $\sigma_{\rm 0}$. The mean time between direct encounters between different types of objects in the cluster core, where the interaction rate dominates and we expect most simulated collisions to occur, can then be expressed using the mean free path approximation:
\begin{equation}
\label{eqn:mfp}
\Gamma = N_{\rm i}n_{\rm j}{\sigma_{\rm i+j}}v_{\rm i+j},
\end{equation} 
where $n_{\rm j}$ is the number density of single stars of type $j$, $v_{\rm i+j}$ is the relative velocity at infinity between the target and the incoming objects, and $\sigma_{\rm i+j}$ is the gravitationally-focused cross-section for collision \citep[e.g.][]{leonard89}:
\begin{equation}
\label{eqn:gf}
\sigma_{\rm i+j} = {\pi}b^2 = {\pi}p^2\Big[1 + \frac{2G(m_{\rm i} + m_{\rm j})}{pv_{\rm i+j}^2}\Big],
\end{equation}
where $b$ is the impact parameter for a pericenter distance $p$ (i.e., distance at closest approach) between two bodies with masses $m_{\rm i}$ and $m_{\rm j}$ that approach each other with a relative velocity at infinity of $v_{\rm i+j}$.  Equation~\ref{eqn:gf} is derived using conservation of energy and (linear and angular) momentum during a two-body encounter.  For the single-single case, we set $p$ $\sim$ ($R_{\rm i} +$ $R_{\rm j}$)  where $R_{\rm i}$ is the radius of particle type $i$.  Equation~\ref{eqn:gf} then becomes \citep{leigh11}:
\begin{eqnarray}
\sigma_{\rm i+j} &=& {\pi}(R_{\rm i} + R_{\rm j})^2\Big[1 + \frac{2G(m_{\rm i} + m_{\rm j})}{(R_{\rm i} + R_{\rm j})v_{\rm i+j}^2}\Big] \nonumber \\
\label{eqn:gf12}
& \approx& \frac{2{\pi}G(m_{\rm i} + m_{\rm j})(R_{\rm i} + R_{\rm j})}{v^2}.
\end{eqnarray}

In order to evaluate whether or not gravitational-focusing should dominate the collisional cross-section, as assumed above, we can compute the Safronov number.  This is the ratio of the escape velocity at the surface of the star to the local root-mean-square velocity, or:
\begin{equation}
\label{eqn:safro}
\Theta_{\rm i} = \frac{{v_{\rm esc,i}^{2}}}{4\sigma^2} = \frac{Gm_{\rm i}}{2\sigma^2R_{\rm i}}.
\end{equation}
If $\Theta_{\rm i} \gg 1$, then gravitational focusing dominates.  In the opposite limit, if $\Theta_{\rm i} \ll 1$, then gravitational focusing is unimportant and the geometric cross-section dominates.  

For our simulations, we have $\sigma$ $=$ 1.06 $\times$ 10$^{-5}$~R$_{\odot}$~s$^{-1}$ and $G =$ 3.96 $\times$ 10$^{-7}$~R$^{3}_{\odot}$~s$^{-2}$~M$^{-1}_{\odot}$, such that $\Theta \approx 175$ for $m_{\rm i}$ $=$ 1 M$_{\odot}$ and  $R_{\rm i}$ $=$ 10~R$_{\odot}$.  For $m_{\rm i}$ and $R_{\rm i}$ with $i =$ (A,B,C,D) in the ranges 1~-~4~M$_{\odot}$ and 1~-~4~R$_{\odot}$, we consistently find $\Theta \gg 1$.  This justifies the approximation made above in Equation~\ref{eqn:gf12}, since gravitational-focusing dominates the collisional cross-section for the range of particle masses and radii considered here. 

We must make one final correction to Equation~\ref{eqn:mfp}, namely a combinatorial correction.  This is done following the procedure or derivation outlined in \citet{leigh18b}.  Specifically, we first convert the number density $n_{\rm j} =$ $N_{\rm j}$/(4$\pi$$r_{\rm c}^3$), where $r_{\rm c}$ is the core radius, and then replace the terms $N_{\rm i}$ and $N_{\rm j}$ with the appropriate combinatorial terms.  For different types of particles we obtain the product $\binom{N_{\rm i}}{1}$$\binom{N_{\rm j}}{1}$, whereas for identical particles we have $\binom{N_{\rm i}}{2}$.

\subsection{Numerical simulations} \label{num}

In this section, we describe the numerical toolkit {\small NBODY6} used in this paper to simulate our N-body systems and to compute the rates of direct collisions between different types of stars.  {\small NBODY6} is a direct N-body simulator that computes the time evolution of clusters by computing pairwise gravitational accelerations.  To reduce the computational cost, this software implements a number of special algorithms such as adaptive time-stepping, the Ahmad Cohen neighbour scheme \citep{Ahmad73}, along with KS \citep{KS65} and Algorithmic \citep{Mikkola99} regularization.  Although accurate, the method comes with significant computational expense, increasing with the total of number of particles roughly as $N^2$.  Some alternative methods exist to soften this scaling, such as the Barnes-Hut tree \citep{BarnesHut86} and particle mesh \citep{Hockney88} algorithms which trade efficiency for accuracy.  

Our simulations initially follow a Plummer \citep{Plummer1911} density distribution.  We consider only single particles initially, and binaries do not form over the extent of our simulations.  We do not consider stellar evolution, such that the initial particle masses and radii remain constant.  

We consider two different suites of simulations, one assuming all identical particles and another adopting different combinations of particle masses and radii.  We set the different combinations of particle masses and radii following \citet{leigh18b}, in order to facilitate a more direct comparison to the results obtained in that paper.  

All of our simulations are performed without external forces, and always begin in a state of virial equilibrium.  Hence, our simulated clusters begin in steady-state and we observe no sudden changes to the cluster properties upon starting our simulations.  In order to generate the required statistics, we perform many repetitions of every simulation, changing only the random seed value.  Specifically, we perform in total 150 simulations for the identical particle case, and in total 360 simulations for those cases with different combinations of the various possible particle types.  

We emphasize that we fix the total particle number for all simulations, specifically to a total of 5000 particles.  This simplification is needed for a controlled experiment aimed at understanding how the theoretical model performs as a function of the particle properties.  In a future study, we will investigate the dependence of the agreement between the simulated and theoretical expectations as a function of the total number of particles.

\begin{figure*}
\begin{center}
\includegraphics[width=\textwidth]{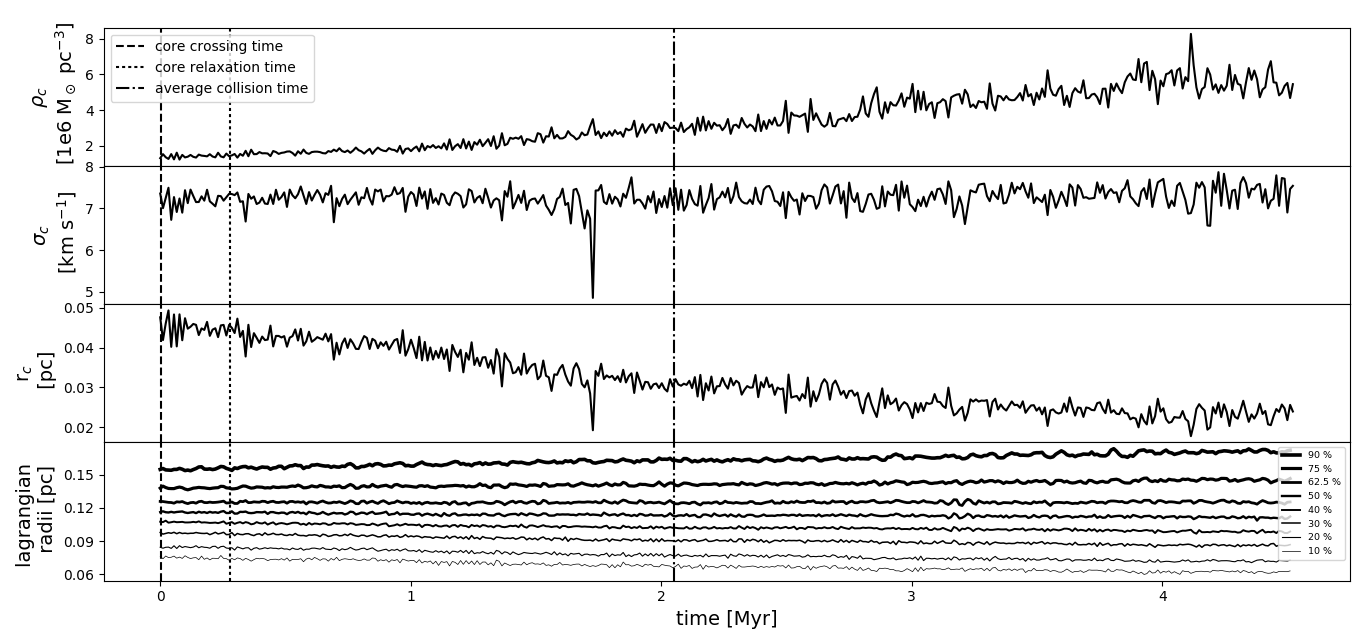}
\end{center}
\caption[Time evolution of the Lagrangian radii, core radius, central velocity dispersion and central density]{The bottom panel shows the time evolution of the 10, 20, 30, 40, 50, 62.5, 75 and 90 percent Lagrangian radii.  Note that the line thickness increases with increasing enclosed mass for the different Lagrangian radii.  The time evolution of the core radius (r$_{c}$), core velocity dispersion ($\sigma_{c}$) and core density ($\rho_{c}$) are also shown in the top three panels.  The dashed line shows the core crossing time, the dotted line shows the core relaxation time and the dash-dotted line shows the average time of first collision. }

\label{fig:fig1}
\end{figure*}

As mentioned before, direct collisions between particles are detected using the ``sticky star approximation".  Here, two stars are assumed to collide (while conserving linear momentum) if their radii overlap in both time and space.  Apart from this, we adopt point particles and neglect for simplicity stellar evolution, tidal interactions and other complicating factors related to the finite-sizes of particles. 
The total runtime is recorded as the time taken until the first collision event, along with the types of particles that collided.  This ensures that our simulations run for at most a few core crossing times, long before the cluster structure has had time to evolve and change significantly due to internal processes such as two-body relaxation.  This is illustrated in Fig.~\ref{fig:fig1}, which shows the time evolution of the core properties over the duration of a particular realization of our simulations for the identical particle case.  

Thus, the assumption that the chosen initial cluster conditions remain
approximately static is valid over the entire extent of our simulations.  This
ensures a proper and robust comparison to our analytic predictions, which
assume single time-averaged values for the cluster properties (radius or size,
velocity dispersion and density).  Importantly, we terminate all simulations
at the time of the first collision.  We prefer this alternative to running
simulations for longer, and recording all times at which different collision
events occur in the same simulation.  This is done to minimize changes to the
host cluster properties over the time our simulations are run (i.e., up until
the time of the first collision event), which ultimately facilitates a more
honest comparison between the simulated and predicted collision times. With
this we ensure that we are comparing simulations with host cluster properties that actually agree to within a few percent with our assumed average cluster properties used to compute the analytic rates.  The next simulation can then be immediately performed, avoiding any additional computational overhead (relative to performing fewer longer simulations). 

\subsection{Identical particles}

We begin with 150 simulations assuming identical particles, with masses and radii of, respectively, 1~M$_{\odot}$ and 10~R$_{\odot}$.  This simple set of initial conditions allows for a baseline comparison between our simulated and predicted collision rates.  We adopt $N = 5000$ particles initially for all sets of simulations, and set the core radius, density and velocity dispersion to be, respectively, 0.05~pc, 1.3~$\times$~10$^6$~M$_{\odot}$~pc$^{-3}$ and 7.4~km~s$^{-1}$. Each of the 150 simulations has the same initial conditions, varying only the random seed value.

See Table \ref{table:tab1} for a summary.

\begin{table}
\begin{tabular}{p{4cm}p{2cm}}
 \hline
 \multicolumn{2}{c}{Initial Conditions} \\
 \hline
 Number of simulations& 150 \\
 Number of particles& 5000 \\
 Particle mass& 1 M$_{\odot}$ \\
 Particle radius& 10 R$_{\odot}$ \\
 Core radius& 0.05 pc \\
 Half-mass radius& 0.11 pc\\
 Core density& 1.3 $\times$ 10$^{6}$ pc$^{-3}$\\
 Core velocity dispersion& 7.4 km~s$^{-1}$ \\
 Core mass& 6.0 $\times$ 10$^{2}$ M$_{\odot}$ \\
 Core crossing time & 6,612 yr \\
 Core relaxation time & 0.3 Myr \\
 \hline
\end{tabular}
\caption{Initial conditions for simulations with identical particles.}
\label{table:tab1}
\end{table}

\subsection{Multiple types of particles}
\label{multi1}

One of the main goals of this paper is to construct a Collision Rate Diagram that is analogous to the one shown in fig.~4 in \citet{leigh18b}, which already includes gravitational-focusing and incorporates the required corrections from combinatorics.  This is done to evaluate the efficacy of the mean free path approximation and the model presented in \citet{leigh17} by facilitating a more direct comparison of the accuracy of the CRD in capturing the dominant physics in both the small- and large-number limits.  
To this end, we perform additional numerical experiments designed to be in close analogy with the procedure outlined in \citet{leigh18b}.  That is, we adopt the same combinations of particle masses and radii, differing relative to \citet{leigh18b} only in the total number of particles but not in the relative fractions of the different particle types.  Specifically, we consider four different types of particles, with masses of 1, 2, 3 and 4~M$_{\odot}$ and corresponding radii of, respectively, 1, 2, 3 and 4~R$_{\odot}$. We call these types of particles A, B, C and D, respectively.  

The initial conditions and our choices for particle combinations are summarized in Table \ref{table:tab2}. For each suite of simulations with the same initial conditions, we show the numbers of simulations performed, the number of particles of every type (A, B, C and D), in addition to the core mass in M$_{\odot}$, the core density in M$_{\odot}$~pc$^{-3}$, the core velocity dispersion in km~s$^{-1}$ and the core radius in pc.  
The above parameters are shown for every one of the twelve different combinations of particle types considered here.  For every such combination, we adopt in total 5000 particles and repeat the simulation 30 times varying only the random seed to help ensure that, upon averaging over all identical simulation sets, the required statistical significance and convergence of our results are achieved.

\begin{table*}
\begin{tabular}{p{0.8cm}p{1.2cm}p{1.2cm}p{1.2cm}p{1.2cm}p{1.3cm}p{1.6cm}p{1.8cm}}
 \hline
 \multicolumn{8}{c}{Initial Conditions} \\
 \hline
 \# &Number particles type A\newline& Number particles type B& Number particles type C & Number particles type D& Core Mass (M$_{\odot}$)& Core\newline Density\newline (M$_{\odot}$ pc$^{-3}$)  & Core Velocity Dispersion (km s$^{-1}$)\\
 \hline 
 1 &2500&0   &1250&1250&1.2 $\times$ 10$^3$  &3.0 $\times$ 10$^6$ &10.8\\
 2&2500&1250&0   &1250&1.1 $\times$ 10$^3$  &2.7 $\times$ 10$^6$&10.2\\
 3&2500&1250&1250&0   &1.0 $\times$ 10$^3$    &2.4 $\times$ 10$^6$&9.6\\
 4&1250&2500&1250&0   &1.2 $\times$ 10$^3$  &2.7 $\times$ 10$^6$&10.5\\
 5 &1250&2500&0   &1250&1.3 $\times$ 10$^3$ &3.1 $\times$ 10$^6$&10.9\\
 6&0   &2500&1250&1250&1.6 $\times$ 10$^3$ &3.7 $\times$ 10$^6$&12.9\\
 7&0   &1250&2500&1250&1.7 $\times$ 10$^3$  &4.0 $\times$ 10$^6$ &12.7\\
 8&1250&1250&2500&0   &1.3 $\times$ 10$^3$  &3,1 $\times$ 10$^6$&11\\
 9&1250&0   &2500&1250&1.6 $\times$ 10$^3$ &3.8 $\times$ 10$^6$&12.2\\
 10&0   &1250&1250&2500&1.9 $\times$ 10$^3$ &4.4 $\times$ 10$^6$&13.3\\
 11&1250&0   &1250&2500&1.7 $\times$ 10$^3$ &4.1 $\times$ 10$^6$&12.6\\
 12&1250&1250&0   &2500&1.6 $\times$ 10$^3$ &3.7 $\times$ 10$^6$&12.1\\
 
  \hline
 
\end{tabular}
\caption{Initial conditions for all simulations with different types of particles.}
\label{table:tab2}
\end{table*}

\section{Results} \label{results}

In this section, we present the results of our numerical N-body simulations and compare them to our analytic predictions from the mean free path approximation.

\subsection{Identical particles} \label{simple2}

We begin with the identical stars case, where all stars in the cluster are initially assumed to have masses and radii of, respectively, 1~M$_{\odot}$ and 10~R$_{\odot}$. We compute \textit{core} collision times as the timescale for single-single stellar encounters following equation A9 from \cite{leigh11} using the values of radius, density and velocity dispersion for the cluster core presented in Table~\ref{table:tab1}. This is because we expect most collisions in our simulations to occur in the core (or very near to it).  Indeed, this is observed in our simulations, but we will return to this assumption later on.

\begin{figure}
\begin{center}
\includegraphics[width=\columnwidth]{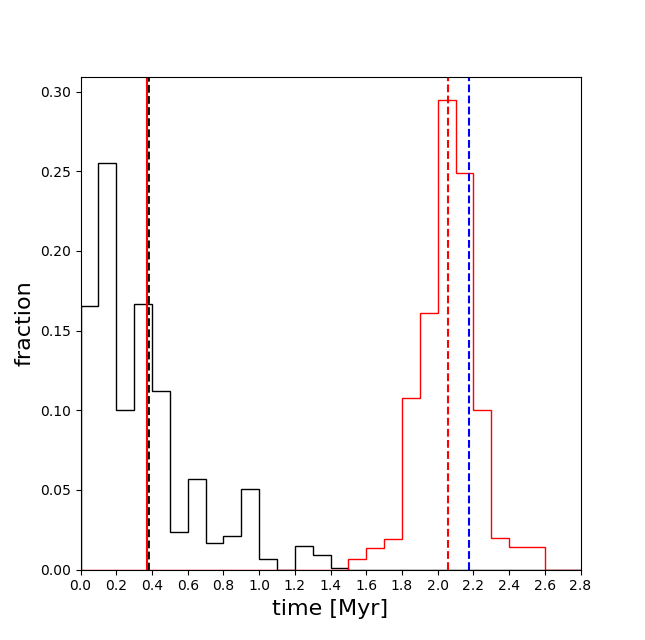}
\end{center}
\caption[Comparing the simulated and analytic distributions of collision times for the identical particle case.]{The black histogram shows the distribution of simulated collision times for the first collision event in all 150 simulations, for the identical particle case.  The red histogram shows the analytically predicted distribution for the \textit{core} collision times using the cluster conditions at the time of the first collision 
in each simulation. The dashed black and red vertical lines indicate the means of each distribution. The blue dashed line shows the average collision time in the \textit{core} using the initial (t $=$ 0) cluster conditions to compute the timescales.  Finally, the solid red vertical line shows the analytic mean collision time using the total integrated collision times over the entire cluster and not just the core \citep{freitag02}.    
Notice that, when using the core collision times, the simulated and predicted results typically agree to within a factor $\lesssim$ 5.  However, when using the total integrated collision times, this factor is roughly corrected for and we see good agreement.}
\label{fig:fig2}
\end{figure}

The results of this comparison are shown in Fig.~\ref{fig:fig2}, where we compare our simulated and analytic predictions for the collision times.  The solid black and red histograms show, respectively, the  simulated results and our theoretical predictions.  The latter are computed using the \textit{core} collision times adopting the cluster conditions at the moment of the first collision in the corresponding simulation. For comparison, we also calculate the \textit{core} collision times adopting the core conditions at the beginning of the simulations (i.e., t$=$0).  The dashed black and red vertical lines show the means of each distribution, with the color of the lines corresponding to the color of the corresponding histogram.  The blue dashed line shows the average collision time in the \textit{core} using the initial (t $=$ 0) cluster conditions to compute the timescales.  Finally, the solid red vertical line shows the mean collision time computed by integrating over the entire extent of the cluster (i.e., out to the tidal radius) \citep{freitag02}. 

Note that the simulated and predicted results typically agree to within a factor $\lesssim$ 5 when using the core collision rates.  When using the total integrated collision rates, this factor is reduced to $\sim$ 1.  Therefore, even though the dominant contribution to the total number of collisions in our simulations comes from the core, the total integrated rates are in much better agreement with those obtained from our simulations. This improvement is not surprising given that not \textit{all} collisions occur in the core in our simulations.    
In the subsequent sections, our focus is on comparing \textit{relative} collision rates.  Hence, we continue to use the same \textit{core} collision rates as before, since integrating over the entire cluster will not change the \textit{relative} rates.  This would only contribute a constant correction factor, which we have shown is of order $\sim$ 5, increasing the overall rates relative to the core rates.  We will return to the issue of comparing the simulated and predicted absolute timescales in Section~\ref{discussion}.

\subsection{Multiple types of particles} \label{four}

Next, we move on to consider simulations involving multiple particle types, as described in Section~\ref{multi1}.  The results of these experiments are summarized in Figs~\ref{fig:fig3},~\ref{fig:fig5} and~\ref{fig:fig6}, as well as Table~\ref{tab:tabl3}. We begin by comparing the simulated and predicted collision time distributions, then move on to using the CRD in Fig.~\ref{fig:fig6} to better quantify the quality of the agreement between our simulations and analytic predictions. As already discussed, we adopt the \textit{core} collision times throughout this section, since we are most interested in the \textit{relative} rates for different types of collisions to occur (e.g., A+A, A+B, B+C, etc.).  As shown in the previous sections, the absolute timescales should be corrected by a factor of order 5, but we emphasize again that this correction factor does not affect our comparisons of the \textit{relative} rates.

\subsubsection{Comparing the distributions of initial collision times} \label{first}

\begin{figure*}
\begin{center}
\includegraphics[width=\textwidth]{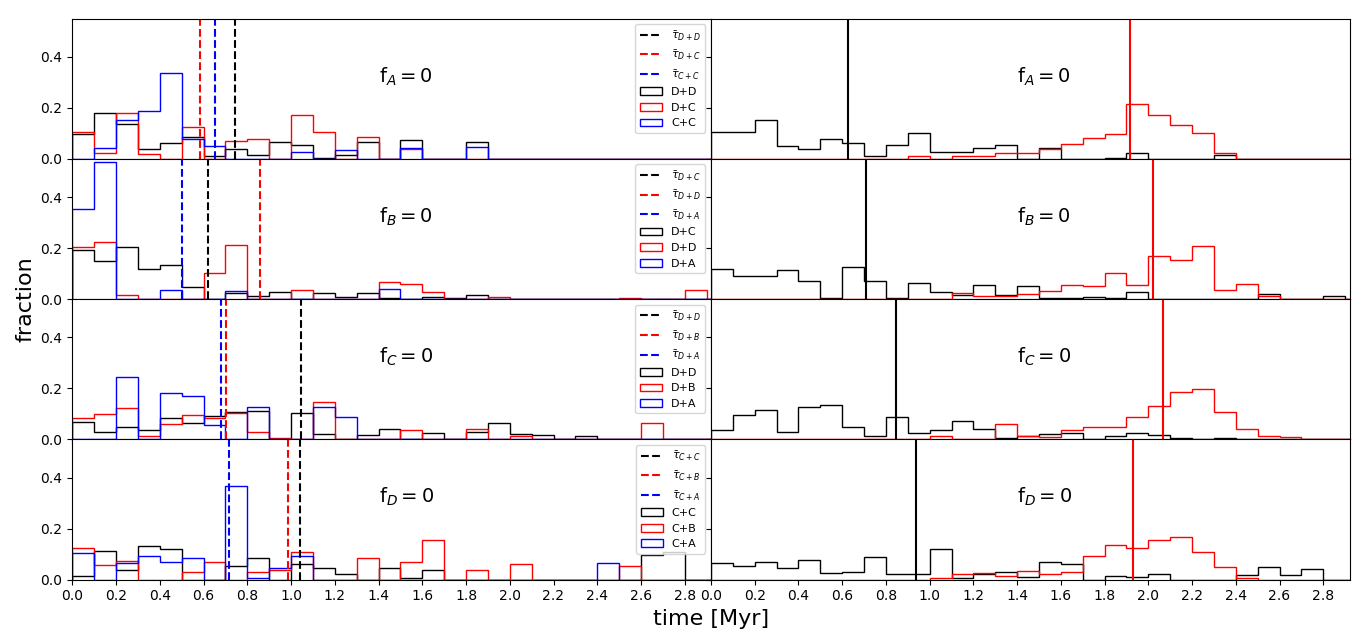}
\end{center}
\caption[Histograms of the collision times for the 3 most frequent types of collisions in our simulations.]{In the left-hand panels, we show histograms of the collision times for the 3 most frequent types of collisions in our simulations, plotted as fractions of the total. Each histogram has a sample size of 30.  The histograms are color-coded, with the black, red and blue histograms corresponding to, respectively, the first, second and third most frequent types of collisions in our simulations.  The dashed vertical lines indicate the means of each distribution. In the right-hand panels, we show the total distribution of collision times for \textit{all} types of collisions combined.  That is, we combine all three rows in each quadrant of Table~\ref{tab:tabl3}, and plot this final distribution for all 90 simulations. The black and red histograms show, respectively, the simulated and predicted distributions of \textit{core} collision times. The black and red vertical lines show the corresponding mean values in any histogram. All analytic predictions are calculated using the core properties at the time of collision.}
\label{fig:fig3}
\end{figure*}

\begin{figure}
\begin{center}
\includegraphics[width=\columnwidth]{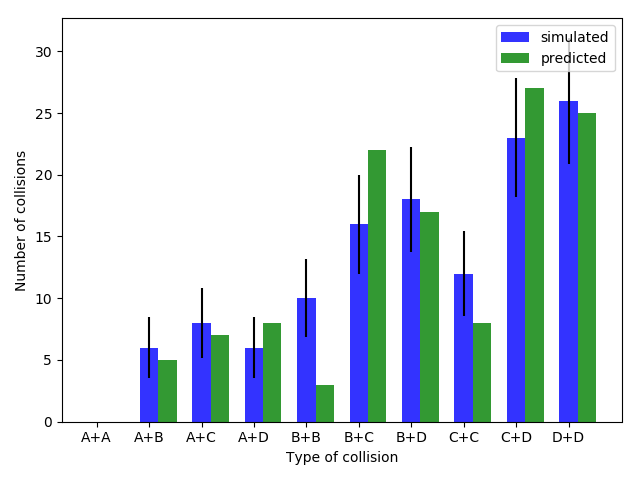}
\label{fig:hist}
\end{center}
\caption[Histograms comparing the number of simulated and predicted collisions for each type (e.g., A+A, A+B, etc.).]{Histograms comparing the number of simulated and predicted collisions for each type (e.g., A+A, A+B, etc.), and including error bars calculated assuming Poissonian statistics.}
\label{fig:fig5}
\end{figure}

In close analogy with Table 1 in \citet{leigh16b}, we summarize the results of our numerical scattering simulations for the multi-particle case in Table~\ref{tab:tabl3}, including the numbers of different types of collisions.  The results are summarized in histogram form in Fig.~\ref{fig:fig3}. Note that the simulated and predicted distributions of collision times are typically consistent with each other to within a factor of $\sim 4$.  The correction is slightly lower than for the identical particle case, due to the larger collisional cross-sections.  This is made more evident by comparing the means of each distribution, indicated by the vertical lines.  However, we also see significantly more scatter in the simulated data relative to our analytic predictions.  This is because we are dealing with a chaotic problem, such that even collision events predicted to be rare or non-existent in the simulations can still occur (though very infrequently, if they occur at all).  This contributes to broadening the overall distribution of collision times in the simulated data set.  Moreover, the analytic predictions all use very similar, albeit slightly perturbed, initial conditions.  For this reason, they tend to predict similar collision times, contributing to more compact (i.e., having a smaller range of collision times) distributions relative to the simulated data sets.

\begin{table*}
\begin{center}
\begin{tabular}{p{1.2cm}p{1.4cm}p{1.4cm}p{1.2cm}p{1.2cm}p{1.4cm}p{1.4cm}p{1.2cm}}
\hline

\hline
\multicolumn{4}{c}{TOP LEFT} 
& \multicolumn{4}{c}{TOP RIGHT} \\ 

\hline

\hline
Type of Collision & Number of Simulated Collisions& Number of Predicted Collisions & Collision Symbols & Type of Collision & Number of Simulated Collisions & Number of Predicted Collisions & Collision Symbol\\ 
\hline
& \multicolumn{2}{c}{(2,0,1,1)}&&&\multicolumn{2}{c}{(2,1,1,0)}&\\
\hline

D+C & 14 & 12 &\begin{tikzpicture}\node[draw=black,minimum size=0.07cm,regular polygon,regular polygon sides=5] (a) {};\end{tikzpicture} &
C+C & 12 & 5 &\begin{tikzpicture}\node[draw=black,minimum size=0.07cm,regular polygon,regular polygon sides=7] (a) {};\end{tikzpicture} \\

D+D & 10 &  5&\begin{tikzpicture}\node[draw=black,minimum size=0.07cm,regular polygon,regular polygon sides=6] (a) {};\end{tikzpicture}&
C+A & 7 &  10 &$\triangle$ \\

C+A & 4  & 4 &$\triangle$&
C+B & 6 & 11 &\begin{tikzpicture}\node[draw=black,minimum size=0.07cm,regular polygon,regular polygon sides=4] (a) {};\end{tikzpicture}\\

D+A & 2  & 8 & $\times$ &
A+A & 2 & 0 & \\

C+C & 0  & 2 &\begin{tikzpicture}\node[draw=black,minimum size=0.07cm,regular polygon,regular polygon sides=7] (a) {};\end{tikzpicture} &
B+A & 2 & 4 & \\

A+A & 0  & 0 & &
B+B & 1 & 1 & \\

 \hline
& \multicolumn{2}{c}{\textcolor{red}{(1,0,2,1)}}&&&\multicolumn{2}{c}{\textcolor{red}{(1,2,1,0)}}&\\

\hline

\textcolor{red}{D+C} & \textcolor{red}{17} & \textcolor{red}{18} &\begin{tikzpicture}\node[draw=red,minimum size=0.07cm,regular polygon,regular polygon sides=5] (a) {};\end{tikzpicture}&
\textcolor{red}{C+B} & \textcolor{red}{13} & \textcolor{red}{17} &\begin{tikzpicture}\node[draw=red,minimum size=0.07cm,regular polygon,regular polygon sides=4] (a) {};\end{tikzpicture} \\
\textcolor{red}{D+D} & \textcolor{red}{4}  & \textcolor{red}{4} &\begin{tikzpicture}\node[draw=red,minimum size=0.07cm,regular polygon,regular polygon sides=6] (a) {};\end{tikzpicture}&
\textcolor{red}{C+C} & \textcolor{red}{5}  & \textcolor{red}{4} &\begin{tikzpicture}\node[draw=red,minimum size=0.07cm,regular polygon,regular polygon sides=7] (a) {};\end{tikzpicture}\\
\textcolor{red}{C+C} & \textcolor{red}{4}  & \textcolor{red}{3} &\begin{tikzpicture}\node[draw=red,minimum size=0.07cm,regular polygon,regular polygon sides=7] (a) {};\end{tikzpicture}&
\textcolor{red}{C+A} & \textcolor{red}{5}  & \textcolor{red}{4} &\textcolor{red}{$\triangle$} \\
\textcolor{red}{C+A} & \textcolor{red}{3}  & \textcolor{red}{3} &\textcolor{red}{$\triangle$} &
\textcolor{red}{B+A} & \textcolor{red}{4}  & \textcolor{red}{3} & \\
\textcolor{red}{D+A} & \textcolor{red}{2}  & \textcolor{red}{3} &\textcolor{red}{$\times$} &
\textcolor{red}{B+B} & \textcolor{red}{3}  & \textcolor{red}{2} & \\
\textcolor{red}{A+A} & \textcolor{red}{0}  & \textcolor{red}{0} & &
\textcolor{red}{A+A} & \textcolor{red}{0}  & \textcolor{red}{0} & \\
\hline
& \multicolumn{2}{c}{(1,0,1,2)}&&&\multicolumn{2}{c}{(1,1,2,0)}&\\
\hline 
D+D & 13 & 7 &\begin{tikzpicture}\node[draw=black,minimum size=0.07cm,regular polygon,regular polygon sides=6] (a) {};\end{tikzpicture}&
C+C & 13 & 7 &\begin{tikzpicture}\node[draw=black,minimum size=0.07cm,regular polygon,regular polygon sides=7] (a) {};\end{tikzpicture}\\

D+C & 7  & 16 &\begin{tikzpicture}\node[draw=black,minimum size=0.07cm,regular polygon,regular polygon sides=5] (a) {};\end{tikzpicture}&
C+B & 11 & 14 &\begin{tikzpicture}\node[draw=black,minimum size=0.07cm,regular polygon,regular polygon sides=4] (a) {};\end{tikzpicture}\\

D+A & 6  & 5 &$\times$&
C+A & 4  & 7 &$\triangle$\\
C+C & 2  & 1 &\begin{tikzpicture}\node[draw=black,minimum size=0.07cm,regular polygon,regular polygon sides=7] (a) {};\end{tikzpicture}&
B+B & 2  & 1 & \\
C+A & 2  & 1 &$\triangle$ &
B+A & 0  & 1 & \\
A+A & 0  & 0 & &
A+A & 0  & 0 & \\

\hline

\hline
\multicolumn{4}{c}{BOTTOM LEFT} & \multicolumn{4}{c}{BOTTOM RIGHT}\\
\hline

\hline
Type of Collision & Number of Simulated Collisions& Number of Predicted Collisions & Collision Symbols & Type of Collision & Number of Simulated Collisions & Number of Predicted Collisions & Collision Symbol\\ 
\hline
& \multicolumn{2}{c}{(2,1,0,1)}&&&\multicolumn{2}{c}{(0,1,2,1)}&\\
\hline 
D+D & 16 & 7 &\begin{tikzpicture}\node[draw=black,minimum size=0.07cm,regular polygon,regular polygon sides=6] (a) {};\end{tikzpicture}&
D+C & 9 & 15 &\begin{tikzpicture}\node[draw=black,minimum size=0.07cm,regular polygon,regular polygon sides=5] (a) {};\end{tikzpicture}\\

D+A & 7 & 11 &$\times$&
C+C & 9 & 2 &\begin{tikzpicture}\node[draw=black,minimum size=0.07cm,regular polygon,regular polygon sides=7] (a) {};\end{tikzpicture}\\

D+B & 4  & 10 &\begin{tikzpicture}
     \draw node[draw=black,circle,minimum size=0.1cm] {};
\end{tikzpicture} &
D+D & 6 & 3 &\begin{tikzpicture}\node[draw=black,minimum size=0.07cm,regular polygon,regular polygon sides=6] (a) {};\end{tikzpicture}\\

B+A & 2  & 2 & &
C+B & 4 & 5 &\begin{tikzpicture}\node[draw=black,minimum size=0.07cm,regular polygon,regular polygon sides=4] (a) {};\end{tikzpicture}\\

B+B & 1 & 1 & &
D+B & 2 & 5 &\begin{tikzpicture}
     \draw node[draw=black,circle,minimum size=0.1cm] {};
\end{tikzpicture} \\

A+A & 0  & 0 & &
B+B & 0 & 0 & \\
\hline
& \multicolumn{2}{c}{\textcolor{red}{(1,2,0,1)}}&&&\multicolumn{2}{c}{\textcolor{red}{(0,2,1,1)}}&\\
\hline 

\textcolor{red}{D+B} & \textcolor{red}{11} & \textcolor{red}{17} &\begin{tikzpicture}
     \draw node[draw=red,circle,minimum size=0.1cm] {};
\end{tikzpicture} &
\textcolor{red}{D+D} & \textcolor{red}{11} & \textcolor{red}{4} &\begin{tikzpicture}\node[draw=red,minimum size=0.07cm,regular polygon,regular polygon sides=6] (a) {};\end{tikzpicture}\\

\textcolor{red}{B+B} & \textcolor{red}{7}  & \textcolor{red}{1} & &
\textcolor{red}{D+B} & \textcolor{red}{7} & \textcolor{red}{10} &\begin{tikzpicture}
     \draw node[draw=red,circle,minimum size=0.1cm] {};
\end{tikzpicture}\\

\textcolor{red}{D+D} & \textcolor{red}{6}  & \textcolor{red}{6} &\begin{tikzpicture}\node[draw=red,minimum size=0.07cm,regular polygon,regular polygon sides=6] (a) {};\end{tikzpicture}&
\textcolor{red}{D+C} & \textcolor{red}{6}  & \textcolor{red}{9} &\begin{tikzpicture}\node[draw=red,minimum size=0.07cm,regular polygon,regular polygon sides=5] (a) {};\end{tikzpicture}\\

\textcolor{red}{D+A} & \textcolor{red}{4} & \textcolor{red}{5} &\textcolor{red}{$\times$} &
\textcolor{red}{C+B} & \textcolor{red}{3}  & \textcolor{red}{5} &\begin{tikzpicture}\node[draw=red,minimum size=0.07cm,regular polygon,regular polygon sides=4] (a) {};\end{tikzpicture}\\

\textcolor{red}{B+A} & \textcolor{red}{2}  & \textcolor{red}{2} & &
\textcolor{red}{C+C} & \textcolor{red}{3} & \textcolor{red}{1} &\begin{tikzpicture}\node[draw=red,minimum size=0.07cm,regular polygon,regular polygon sides=7] (a) {};\end{tikzpicture}\\

\textcolor{red}{A+A} & \textcolor{red}{0} & \textcolor{red}{0} & &
\textcolor{red}{B+B} & \textcolor{red}{0}  & \textcolor{red}{0} & \\

\hline
& \multicolumn{2}{c}{(1,1,0,2)}&&&\multicolumn{2}{c}{(0,1,1,2)}&\\
\hline
D+D & 14 & 9 &\begin{tikzpicture}\node[draw=black,minimum size=0.07cm,regular polygon,regular polygon sides=6] (a) {};\end{tikzpicture}&
D+D & 16 & 5 &\begin{tikzpicture}\node[draw=black,minimum size=0.07cm,regular polygon,regular polygon sides=6] (a) {};\end{tikzpicture}\\
D+B & 11 & 13 &\begin{tikzpicture}
     \draw node[draw=black,circle,minimum size=0.1cm] {};
\end{tikzpicture} &
D+C & 7  & 14 &\begin{tikzpicture}\node[draw=black,minimum size=0.07cm,regular polygon,regular polygon sides=5] (a) {};\end{tikzpicture}\\
D+A & 3  & 7 &$\times$&
C+C & 4  & 1 &\begin{tikzpicture}\node[draw=black,minimum size=0.07cm,regular polygon,regular polygon sides=7] (a) {};\end{tikzpicture}\\
B+B & 1 & 0 & &
D+B & 3 & 8 &\begin{tikzpicture}
     \draw node[draw=black,circle,minimum size=0.1cm] {};
\end{tikzpicture} \\
A+A & 1 & 0 & &
B+B & 0 & 0 & \\
A+B & 0  & 1 & &
B+C & 0  & 2 &\begin{tikzpicture}\node[draw=black,minimum size=0.07cm,regular polygon,regular polygon sides=4] (a) {};\end{tikzpicture}\\
\hline 

 \hline
 
\end{tabular}
\caption{Comparing the simulated and predicted numbers of collisions, for different combinations of particle types.  We highlight those insets in red that were used to construct the CRD in Figure~\ref{fig:fig6}. }
\label{tab:tabl3}
\end{center}
\end{table*}

\subsubsection{Comparing the analytic and simulated rates via collision rate diagrams}

\begin{figure*}
\begin{center}
\includegraphics[width=12 cm]{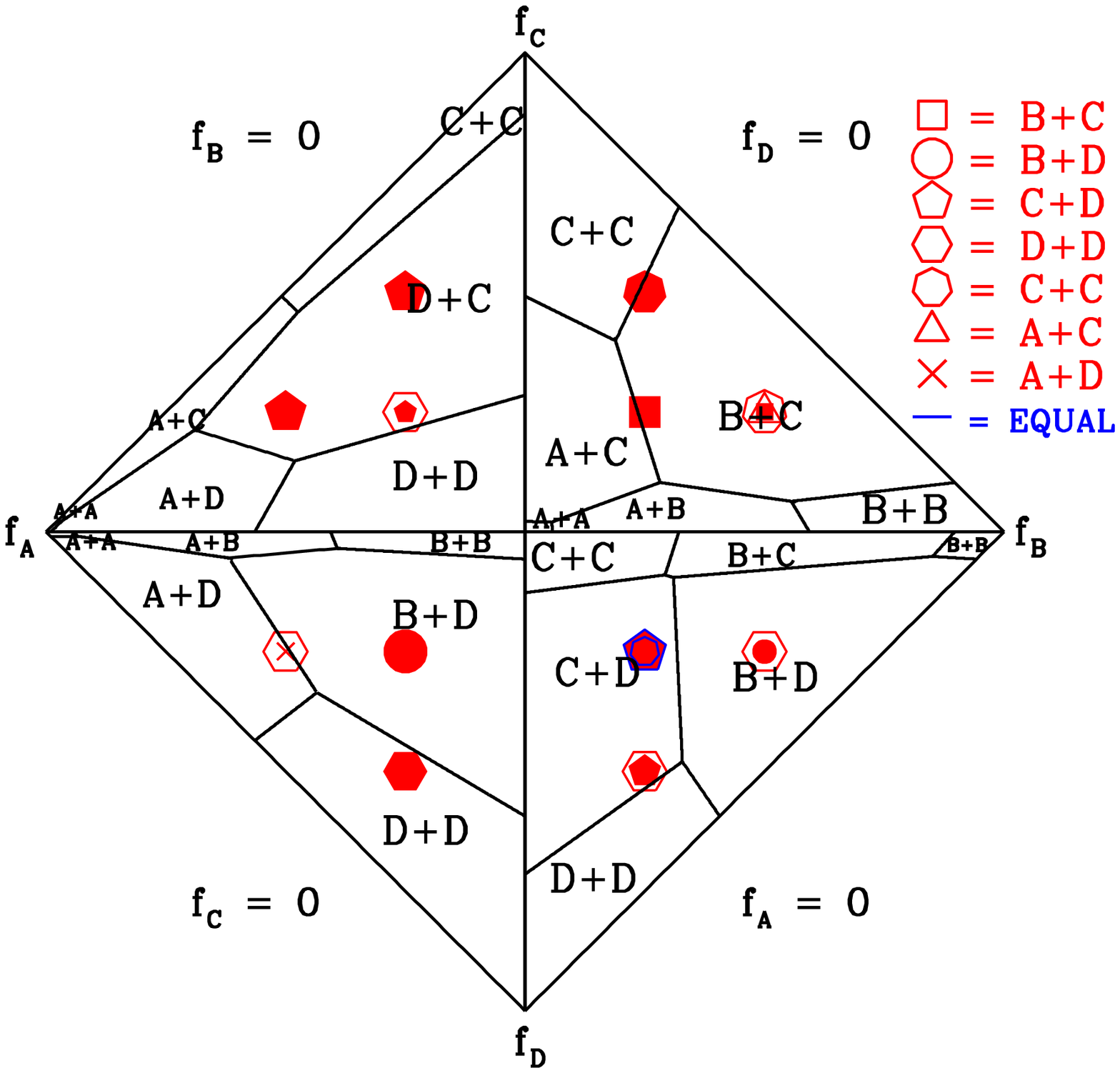}
\label{fig:crdn}
\end{center}
\caption[CRD adopting a combinatorial correction with the data from our \texttt{NBODY6} simulations over-plotted.]{Collision Rate Diagram adopting both gravitational-focusing and the appropriate combinatorial correction, with the simulated data over-plotted.  Each quadrant shows the parameter space in the f$_{\rm i}$-f$_{\rm j}$-plane for which different types of collisions dominate.  The types of collisions associated with each type of symbol are shown in the upper right portion of the figure.  Closed symbols indicate perfect agreement between the simulations and the prediction of our CRD.  Open symbols indicate disagreement.  In these cases, we plot the dominant type of collision (i.e., that happens the most) using the largest open symbol.  We then plot the second most frequent type of collision within that symbol, but slightly smaller.  We carry on in this way for the third, fourth, etc. most frequent types, until a filled symbol is eventually plotted.}
\label{fig:fig6}
\end{figure*}

The results of our collision statistics for the \textit{relative} rates of collisions can be summarized via a Collision Rate Diagram, as shown in Fig.~\ref{fig:fig6}.  Closed symbols indicate perfect agreement between the simulations and the prediction of our CRD.  Open symbols indicate disagreement.  The key point to take away is that most points are closed symbols, indicating that the simulated and predicted results are in agreement.  Moreover, where we do see open symbols, a quick glance at Table~\ref{tab:tabl3} shows that the disagreement is minor, and can likely be attributed to low-number statistics, and the overall effects of chaos and the subsequent introduction of statistical fluctuations. 

The effects of low-number statistics and chaos, in addition to relevant physical processes that are not included in our model might be contributing to the few simulation sets for which the agreement is poorest as illustarted in
Fig.~\ref{fig:fig5}. Here, we plot for each type of collision event (i.e., A+A, A+B, etc.) the total number of collisions obtained in all 90 simulation sets, for the combinations of particle types adopted in Fig.~\ref{fig:fig6}, along with the 1-$\sigma$ uncertainties calculated assuming Poissonian statistics.  For comparison, we also show the predictions from our analytic calculations.  As is clear, most of our simulated results agree with our analytic predictions to within one standard deviation.

\section{Discussion} \label{discussion}

In this section, we begin by evaluating the efficiency of the mean free path approximation.  This is done by first discussing the agreement obtained between our theoretical model and the simulations, and then quantifying the sources of the slight discrepancies we do find in our results.  We then go on to discuss possible improvements to our base model, and how to apply the mean free path approximation in more complicated time-dependent gravitational potentials than considered here (e.g., including two-body relaxation and mass segregation, stellar evolution, and so on).

\subsection{Quantifying the agreement between the theoretical predictions and the simulations}

Our analytic predictions typically agree with the simulated predictions to within one standard deviation for the \textit{relative} rates.  This is illustrated in Figs.~\ref{fig:fig2}-\ref{fig:fig6}, and Table~\ref{tab:tabl3}.  We note that the predicted collision times shown in Fig.~\ref{fig:fig3} are longer than the simulated collision times by a factor of a few.  But, as already illustrated in Fig.~\ref{fig:fig2}, we have verified that using the total integrated collision rates for the entire cluster (and not just the core) corrects this discrepancy by roughly the appropriate factor.  Hence, although a non-negligible fraction of collisions occur outside the cluster core in our simulations, the \textit{relative} rates are in good agreement.  This agreement should remain unchanged if we were to use the total integrated collision rates instead, since this only introduces a constant factor which we have already shown is of order $\sim$ 5.  We have also tried relaxing the approximation in Equation~\ref{eqn:gf12} and including the geometric cross-section term in our calculations to see if this improves the agreement.  But, as expected from our preliminary calculations of the Safronov number in Section~\ref{simple}, our results remain unchanged and gravitational focusing clearly dominates over the geometric cross-section.

\begin{figure*}
\begin{center}
\includegraphics[width=\textwidth]{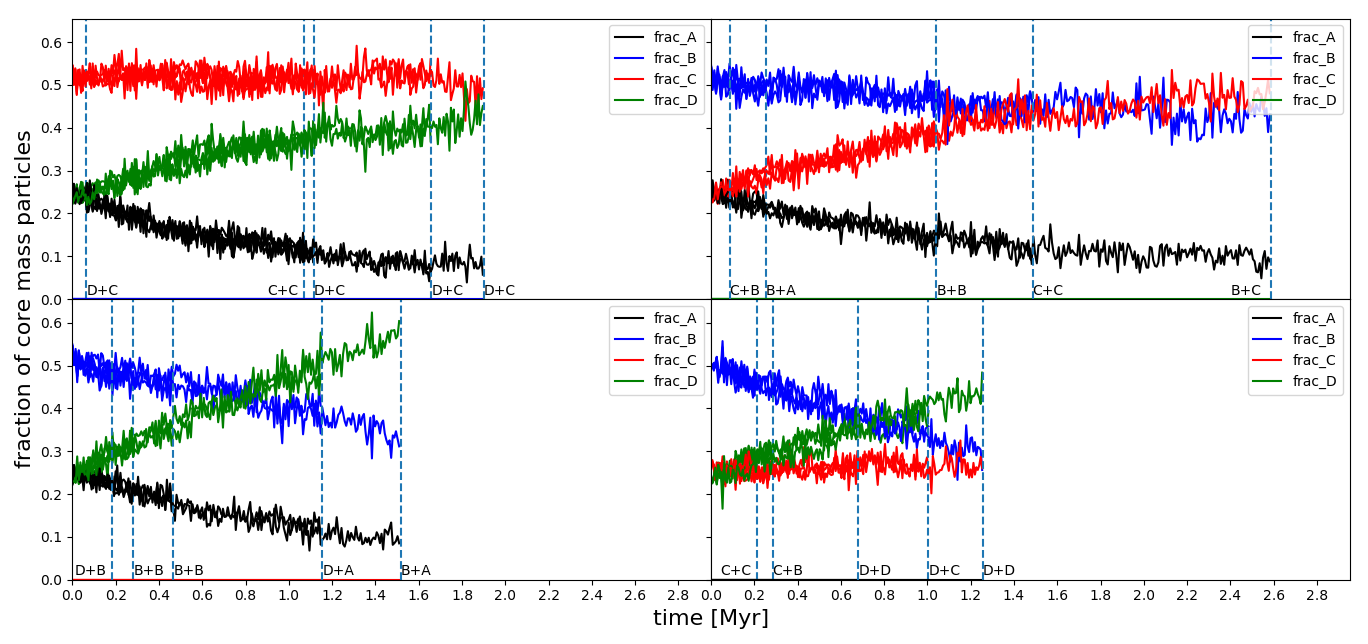}
\end{center}
\caption[Time evolution of the relative numbers of each particle type in the core for every simulation.]{Time evolution of the relative numbers of each particle type in the core for every simulation. In the upper right ($f_{\rm D} = 0$), upper left ($f_{\rm B} = 0$), bottom left ($f_{\rm C} = 0$) and bottom right panels ($f_{\rm A} = 0$) we show the time evolution for the fractions of each particle species.  This time extends from the beginning of the simulations (i.e., t $=$ 0) until the moment of the first collision.  To ensure legibility and adequate presentation of the results, we show the time evolution for 5 (out of 30) simulations in each panel. The fraction of A, B, C, and D particles are represented by, respectively, the black, blue, red and green lines. The vertical dashed lines show the moment when the first collision occurred in that simulation (at which point the simulation was stopped). We include labels to indicate the type of collision that occurred in each simulation.}

\label{fig:fig7}
\end{figure*}

The efficiency and accuracy of our model are illustrated and summarized in the
Collision Rate Diagram shown in Fig.~\ref{fig:fig5}, as well as
Fig.\ref{fig:fig6} and Table~\ref{tab:tabl3}.  Those few simulation sets with
poor agreement between our theoretical predictions and the results of our
simulations can be accounted for primarily due to the effects of mass
segregation, which our model does not include.  This is illustrated in
Fig.~\ref{fig:fig7}, which shows that the fractions of each particle type in
the core can change by a factor of order 2 within a time shorter than it takes
for the first collision to occur.  This overall effect is also illustrated in
Table~\ref{tab:tabl3}, which shows that the agreement between theory and
simulations is typically worse for simulations with more of the heaviest particles (e.g., two D-type particles).  

Secondary effects contributing to the differences between simulations and theory are small-number statistics and chaotic effects.  We have performed a large number of simulations to try to compensate for these effects, but our chosen balance between simulation run-time and number of simulations is still sufficiently small that stochastic effects (e.g., a type of collision that should be very rare happens very early on in the simulation by chance) could still contribute non-negligibly to polluting our final statistics.  Regardless, we have performed enough simulations that these effects should be minor, as quantified by the uncertainties in Fig.~\ref{fig:fig5}.  This issue would be better quantified in a future study using a larger suite of simulations to better explore the issue of convergence.

Independent of the above sources of discrepancy, our model predictions come very close to reproducing the simulated \textit{relative} collision rates for each particle type, in all simulation sets.  This is quantified in Fig.~\ref{fig:fig5}, which illustrates that in most simulation sets the analytic predictions are within 1-$\sigma$ of the simulations.

\subsection{Improving the model for application to more realistic dynamical environments}

Based on our results, we conclude that the mean free path approximation works well for larger-N systems than those considered in \citet{leigh17} and \citet{leigh18b}, provided the particle mass ratios are small (i.e., all particles have similar masses and radii), but worsens for more extreme mass ratios. In our simulations the main reason for this relates to the dependence of the mean collision time on the mean particle mass, which is shorter for higher masses, as illustrated in Figs.~\ref{fig:fig3} and~\ref{fig:fig7}. If collisions are occurring too quickly, it is possible that the assumptions going into our theoretical predictions do not have sufficient time to be reached since we see an increased sensitivity to the initial conditions, specifically the initial positions and velocities of the particles.  Another potential reason for this is that the simple ``n$\sigma$v" approximation for computing the collision rates becomes an over-simplification in the limit of large mass ratios.  For example, the dynamics could be more appropriately treated using a loss-cone approximation.  In future work, this could be incorporated into the model as the mass spectrum becomes more extreme, similar to what was done in \citet{leigh16a} (see, for example, \citet{merritt13} for more details on loss-cone dynamics in the large-N limit).

We also conclude that, at least for the narrow set of initial conditions considered here, the mean free path approximation works better for lower particle numbers relative to larger-N systems, based on the model results presented in \citet{leigh18b}.  This is largely because the time-averaging approximations invoked to construct the model in \citet{leigh18b} begin to break down in the large-N particle limit.  Specifically, the mean timescale for the systems to reach a state of equipartition in the small-number limit is much shorter than the mean collision time, whereas it is the opposite in the large-N limit.  For example, if collisions are occurring, the system is never in a stable steady-state, and the numbers and properties of the particles, as well as the properties of the cluster potential, are continuously changing in time.  Future models should try to account for the effects of a time-dependent cluster potential, time-dependent radial profiles for different mass species, a time-dependent mass function (due not only to collisions, but also evaporation across the tidal boundary of the cluster due to two-body relaxation and also stellar evolution), and so on.  This can be accomplished via a Monte Carlo approach, and breaking the cluster potential up into finite-sized radial shells.  The approach is largely the same as considered here for the cluster core, but applied to each shell independently using an appropriately chosen time-step and summed over all shells.  This is similar to the approach adopted in the leading models for cluster evolution in the literature today \citep{giersz98,giersz01}, and even simpler semi-analytic models and their comparisons to empirical data \citep[e.g.][]{leigh11b,leigh12,leigh15b}.  Further extending and improving upon the model presented in this paper in such a direction will be the focus of future work.  We intend to perform a more detailed parameter space study in future work, keeping the number of particles as a free parameter, with the goal of identifying the approximate particle number at which the assumptions going into the model presented in \citet{leigh17} begin to break down, motivating the switch to a Monte Carlo-based approach.

Finally, several other mechanisms operating rarely could also be contributing to some of the deviations seen in the agreement between our theoretical predictions and the simulations.  For example, binary formation can occur via three-body interactions involving all single stars in high-density environments, although this occurs very rarely in our simulations.  Once formed, these binaries can subsequently merge due to perturbative interactions with single stars causing the binary orbital parameters, in particular the eccentricity, to change in time and eventually lead to a merger.  The model presented in this paper does not account for such complicating effects, although this particular effect has had a negligible impact on our results due to the paucity of binaries formed via three-body interactions in our simulations.  Future work could easily account for this mechanism, however, simply by using a Monte Carlo approach including a rate for binary formation via three-body interactions, combined with a rate of perturbative encounters with single stars and by sampling from the corresponding distributions of orbital parameters \citep[e.g.][]{geller19,hamers19a,hamers19b}.

\section{Summary} \label{summary}
In this paper we test the mean free path approximation by confronting analytic and simulated collision rates in idealized stellar systems, moving to larger particle numbers ($N$~$>$~1000) than in our previous work. We study scenarios with equal mass particles and scenarios including up to four different types (i.e., different masses and radii) of particles modelled following a Plummer distribution and including particle collisions by means of the sticky sphere approximation.

For the equal mass particle case the analytic and simulated collision numbers agree to within a factor of $\sim$5 if only collisions occurring in the core are considered (as is the case for the timescales provided in the Appendix of \citet{leigh11}, for example).  However, in our simulations a dominant fraction of collisions do occur in the core, comparing to the integrated collision rates (i.e., integrated over the entire spatial extent of the cluster and not just the core) corrects for this offset (i.e., corrects by a factor $\sim$ 5 to shorter collision times).  Relative to the identical particle case, we also explored a broader parameter space with simulations that include different combinations of particle types and employ the Collision Rate Diagram to illustrate the results.  We show that we find good agreement between our theoretical predictions and the simulations, finding collision numbers that typically agree to within one standard deviation.

Our results show that the mean free path approximation, and specifically the model presented in \citet{leigh17}, works better for similar particle types (i.e. small mass ratios and size ratios) and for lower particle numbers. The former result is probably caused by the fact that our model does not include the effects of mass segregation and cluster evolution, which are observed in our simulations.  The latter result is possibly related to the longer relaxation times for larger-N systems compared to the corresponding collision times, which causes some of the approximations going into our model (e.g., time-averaged quantities in the low particle number limit) to break down.

In future work, we intend to expand upon and improve the base model considered here for application to large-N systems, by expanding the parameter space considered in this paper to include a range of total particle numbers and cluster properties.  This will be done to identify the critical particle number for which the assumptions going into the model presented in \citet{leigh17} begin to break down.  With this additional step, we will have identified the regions of parameter space where adopting our small-N model from \citet{leigh17} remains valid, and those regions of parameter space where using a shell-based Monte Carlo approach to more accurately model the time evolution of the cluster evolution becomes a better approximation.  In particular, the small-N model from \citet{leigh17} works well for predicting collision probabilities during individual few-body interactions ($N$ = 3, 4, 5, 6, etc.), and could in principle be used to replace small-number gravity integrators called upon in most Monte Carlo-based models for cluster evolution used in the literature today.  Here, a shell-based model for the cluster is used to compute the single-binary and binary-binary interaction rates in each shell, and our formalism could be used to evaluate collision probabilities for the outcomes of these interactions.  However, the critical particle number at which this transition occurs must first be identified, in order to expect the model to match simulated data.  Identifying this critical particle number will be the focus of future work. 

\section{Acknowledgments}
N.W.C.L. gratefully acknowledges support from the Chilean government via Fondecyt Iniciaci\'on Grant \#11180005. 
B.R. acknowledges funding through Conicyt (CONICYT-PFCHA/Doctorado acuerdo bilateral DAAD/62180013), and DAAD (funding program number 57451854).  DRGS thanks for funding via Conicyt PIA ACT172033 and Fondecyt Regular 1201280.  The authors would also like to sincerely thank Stefano Bovino and Michael Fellhauer for their considerable help accessing and utilizing the computational resources at the Universidad de Concepci\'on.

\section*{Data availability}
The data underlying this article will be shared on request to the corresponding author.

\end{document}